\documentclass[twocolumn,showpacs,aps,prl,superscriptaddress]{revtex4}
\usepackage{graphicx}
\usepackage{dcolumn}
\usepackage{amsmath}
\usepackage{epsfig}
\usepackage{dcolumn}
\usepackage{natbib}
\usepackage{psfrag}

\newcommand{\BaBarType}      {PUB}  
\newcommand{\BaBarYear}       {08}
\newcommand{\BaBarNumber}     {042}
\newcommand{\SLACPubNumber} {13397}

\def\bbar  {\ensuremath{\overline b}\xspace}
\def\Y#1S{\ensuremath{\Upsilon{(#1S)}}\xspace }
\def\OneS  {\Y1S}

\def\ThreeS{\Y3S}
\def\FourS {\Y4S}
\def\FiveS {\Y5S}
\def\SixS {\Y6S}
\def\yfive {\ensuremath{\Upsilon(10860)}\xspace }
\def\ysix {\ensuremath{\Upsilon(11020)}\xspace }

\def\SqrtS {\ensuremath{\sqrt{s}}}

\usepackage{relsize}
\def\babar{\mbox{\slshape B\kern-0.1em{\smaller A}\kern-0.1em
    B\kern-0.1em{\smaller A\kern-0.2em R}}}

\RequirePackage{xspace}

\newcommand{\gev}{\ensuremath{\mathrm{\,Ge\kern -0.1em V}}\xspace}
\newcommand{\mev}{\ensuremath{\mathrm{\,Me\kern -0.1em V}}\xspace}
\def\invpb {\ensuremath{\mbox{\,pb}^{-1}}\xspace}
\def\invfb   {\ensuremath{\mbox{\,fb}^{-1}}\xspace}
\def\epem       {\ensuremath{e^+e^-}\xspace}
\newcommand{\gevc}{\ensuremath{{\mathrm{\,Ge\kern -0.1em V\!/}c}}\xspace}
\newcommand{\mevc}{\ensuremath{{\mathrm{\,Me\kern -0.1em V\!/}c}}\xspace}
\newcommand{\gevcc}{\ensuremath{{\mathrm{\,Ge\kern -0.1em V\!/}c^2}}\xspace}
\newcommand{\mevcc}{\ensuremath{{\mathrm{\,Me\kern -0.1em V\!/}c^2}}\xspace}
\def\pep2{PEP-II}

\long\def\inst#1{\par\nobreak\kern 4pt\nobreak
    {\it #1}\par\vskip 10pt plus 3pt minus 3pt}

\begin{document}

\begin{flushleft}
\babar-\BaBarType-\BaBarYear/\BaBarNumber \\ 
SLAC-PUB-\SLACPubNumber \\
\end{flushleft}
\begin{flushright}

\end{flushright}

\title{Measurement of the  $e^+e^-\to b\bar{b}$ cross section between
\SqrtS=10.54 and 11.20 \gev}

\begin{abstract}
\noindent
We report  $e^+e^-\to b\bar{b}$ cross section measurements by the \babar\
experiment performed during an energy scan in the range of 10.54 to 11.20 \gev
at the PEP-II $e^+e^-$ collider. 
A total relative error of about 5\% is reached in more than three hundred 
center-of-mass energy steps, separated by  about 5\mev.
These measurements can be used to derive
precise information on the parameters of the \yfive and \ysix resonances. In 
particular we show that their widths may be smaller than previously measured.
 \end{abstract}


%
\author{B.~Aubert}
\author{M.~Bona}
\author{Y.~Karyotakis}
\author{J.~P.~Lees}
\author{V.~Poireau}
\author{E.~Prencipe}
\author{X.~Prudent}
\author{V.~Tisserand}
\affiliation{Laboratoire de Physique des Particules, IN2P3/CNRS et Universit\'e de Savoie, F-74941 Annecy-Le-Vieux, France }
\author{J.~Garra~Tico}
\author{E.~Grauges}
\affiliation{Universitat de Barcelona, Facultat de Fisica, Departament ECM, E-08028 Barcelona, Spain }
% ...
\collaboration{The \babar\ Collaboration}
\noaffiliation

\maketitle

Recent discoveries of  non-baryonic charmonium states that do not behave 
as
two-quark states~\cite{XYZ} call for a search for other resonances 
belonging to this
possible new spectroscopy. Given the charmonium content of these new states, one
could infer the presence of similar resonances containing $b$ quark pairs.
The observed $J^{PC} = 1^{--}$ exotic states ($Y(4260)$, $Y(4350)$, and
$Y(4660)$~\cite{BF-Y}) scaled up by the mass difference between the
$J/\psi$ and the $\OneS$ ($\Delta M \sim 6360 \mevcc$) would be
 exotic bottomonium states with masses above the \FourS\ 
and below 11.2\gev. Moreover, the \yfive and the \ysix states, 
which are candidate \FiveS and \SixS respectively, 
 were observed in the same region~\cite{Y_CLEO,Y_CUSB}.

Between March 28 and  April 7, 2008 
the PEP-II \epem\ collider~\cite{pep2} delivered colliding beams at 
a center-of-mass energy (\SqrtS) in the range of 10.54 to 11.20 \gev. First, an
energy scan over the whole range in 5 \mev steps, collecting approximately
25\invpb per step for a total of about 3.3 \invfb, was performed.
It was then followed by a 600\invpb scan in the range of \SqrtS=10.96 to 11.10
\gev, in 8 steps with non-regular energy spacing, performed in order to
investigate the \SixS region. This data set outclasses the previous
scans~\cite{Y_CLEO,Y_CUSB} by a factor $> 30$ in the luminosity and $\sim 4$ in
the size of the energy steps. Across the scan, the energy of the 
positron beam was kept
fixed at 3.12 \gev, while the electron beam energy was varied accordingly,  to
set the required \SqrtS. This produced a variation of the 
boost of the center-of-mass frame during the scan.

In this Letter we present, for each step in \SqrtS, the 
measurement
 of $ R_b(s) = \sigma_{b}(s)/\sigma_{\mu\mu}^0(s) 
$,
where $\sigma_{\mu\mu}^0 =4\pi\alpha^2/3s$ 
is the lowest-order cross section for $e^+ e^- \to
\mu^+ \mu^-$ and $\sigma_b$ is the total cross section for $e^+ e^-
\to b \bbar (\gamma)$, 
including  $b\bbar$ states produced in initial state radiation ($ISR$)
below the open beauty threshold,
i.e. the $\Y1S$, $\Y2S$, and $\Y3S$ resonances.

The particles produced in the collisions are detected by 
the \babar\ detector, described elsewhere~\cite{babar}. 
Charged-particle tracking is provided by a five-layer silicon
vertex tracker (SVT) and a 40-layer drift chamber (DCH).
In addition to providing precise position information for tracking, 
the SVT and DCH also measure the specific ionization ($dE/dx$), 
which is used for particle identification of low-momentum charged particles. 
At higher momenta ($p>0.7$~\gevc) pions and kaons are identified 
by Cherenkov radiation detected in a ring-imaging device (DIRC). 
The position and energy of neutral clusters (photons) are
measured with an electromagnetic calorimeter (EMC) consisting 
of 6580 thallium-doped CsI crystals.
These systems are mounted inside a 1.5-T solenoidal super-conducting magnet. 
Muon identification is provided by the  magnetic
flux return system instrumented with Resistive Plate Chambers and 
Limited Streamer Tubes.     
The full detector is simulated, for background and efficiency studies, 
with a Monte Carlo program (MC) based on GEANT4~\cite{GEANT4}.

To measure $R_b$, we count the number of events
passing a selection that enriches the sample in events containing 
$B$ mesons ($N_h$) and those passing an
independent 
di-muon selection ($N_\mu$) at each energy point and at a reference energy
below the open beauty production threshold. Indicating with a prime the quantities at the
reference energy, we write:
\begin{eqnarray}
 N_h(s) &=& \Bigg[ \left( R_b(s)\,\sigma^0_{\mu\mu}(s) -
\sigma_{ISR}(s)\right)
 \epsilon_B(s) \\ \nonumber
 &+& \sum_X\sigma_X(s)\,\epsilon_X(s) +
\sigma_{ISR}(s)\,\epsilon_{ISR}(s)
\Bigg] {\cal{L}}(s)\\
N_h^{\prime} &=&
  \left(\sum_X\sigma_X^{\prime}\,\epsilon_X^{\prime}
+ \sigma_{ISR}^{\prime}\,\epsilon_{ISR}^{\prime}\right)
 \,{\cal{L}}^{\prime} \label{eq:nhp}\\
N_\mu(s) &=& \sigma_{\mu\mu}(s)\,\epsilon_\mu(s)\,{\cal{L}}(s)\\
N_\mu^{\prime} &=&
\sigma_{\mu\mu}^{\prime}\,\epsilon_\mu^{\prime}\,{\cal{L}}^{\prime}
\label{eq:method}
\end{eqnarray}
where $\epsilon_B$ is the efficiency for open $b$ production to 
satisfy the hadronic selection, $X$ represents the different background
components described later,  $\sigma_i$ represents 
the cross-sections for the process $i$, $\epsilon_i$
the corresponding efficiency, and ${\cal{L}}$ is the integrated luminosity  
collected at a given value of \SqrtS. Measurements of $N_\mu$ and
$N_\mu^{\prime}$ are needed in order to normalize the hadronic rates to the
collected luminosities.
As reference we choose the sample collected at \SqrtS=10.54\gev,
about 40\mev below the \FourS mass, taken during 2006-2007.
Special mention is made of the 
 $ISR$ sample, the production of
$\Upsilon(nS)$ ($n=1,2,3$) mesons via initial state radiation: albeit 
part of the signal, this 
process can occur at the reference energy and has an efficiency and an
energy dependence of the cross-section different from the open beauty
production.

Solving the system of equations one obtains: 
\begin{equation}
\label{eq:rb}
R_b=\left(\frac{N_h(s)}{N_\mu(s)}-\frac{N_h^{\prime}}{N_\mu^{\prime}}\kappa_{
\sigma\epsilon }(s) \right)\frac{\epsilon_\mu(s)\xi_\mu}{\epsilon_B(s)}\, + R_{ISR}(s),
\end{equation}
where we defined:
\begin{eqnarray}
\nonumber \kappa_{\sigma\epsilon}(s)&=&\frac{\epsilon_\mu^ { \prime } }
{ \epsilon_\mu(s)} \times \\
&\times&\frac{\sum_X R_X(s)\epsilon_X(s) +
R_{ISR}(s)\,\epsilon_{ISR}(s)}{\sum_X
R_X^{\prime}\epsilon_X^{\prime} + R_{ISR}^{\prime}\,\epsilon_{ISR}^{\prime}}
\,,
\end{eqnarray}
and $R_i = \sigma_i/\sigma^0_{\mu\mu}$ for each process and
$\xi_\mu=\sigma_{\mu\mu}/\sigma^0_{\mu\mu}$, assumed independent of 
\SqrtS. 
It should be noted that these equations assume that 
the background scales with the integrated luminosity, i.e. that 
the machine background is negligible, and that the  di-muon 
selection leaves a negligible level of background.

We select the $b$-enriched sample by requiring at least three tracks 
in the event, a total visible energy
in the event greater than 4.5\gev, and a vertex reconstructed from the observed
charged tracks within 5 mm of the beam crossing point in the plane 
transverse to the beam axis and $6\,\rm{cm}$ along the beam axis.
These quantities are computed using exclusively tracks in the 
fiducial volume of the 
DCH
(i.e. forming an angle with the beam axis $0.41<\theta<2.54\,{\rm rad}$).
A further rejection of the main backgrounds,
$e^+e^-\to q\bar{q}$, $q=u,d,s,c$ events 
(``continuum'' events) 
and  $e^+e^-\to \ell^+ \ell^-$, $\ell=e,\mu,\tau$ events, is obtained 
by means of a cut on the ratio of the second and zeroth Fox-Wolfram
moments~\cite{r2}, $R_2$, calculated using only the charged tracks. After 
optimization
of the statistical sensitivity, we require $R_2<0.2$.  
Events that pass this selection at the reference energy comprise
91\% continuum, 2\% two photon ($\epem \to \epem\gamma^*\gamma^*\to\epem X_h$), 
and 
7\% $ISR$
($\epem \to \Upsilon(nS)\gamma_{ISR}$) events.

To select di-muon events, we require that two tracks have an invariant 
mass
greater than 7.5 \gevcc; their angle with the beam 
axis 
in the center-of-mass frame, $\theta_{cms}$, must satisfy
$\cos\theta_{cms}<0.7485$, and the two muons must be collinear to within
10$^o$. 
To exploit the fact that muons are minimum ionizing particles, we
require that at least one of them leaves a signal in the EMC, and neither 
deposits more than 1 \gev.
\begin{figure*}[htb]
\begin{center}
\epsfig{file=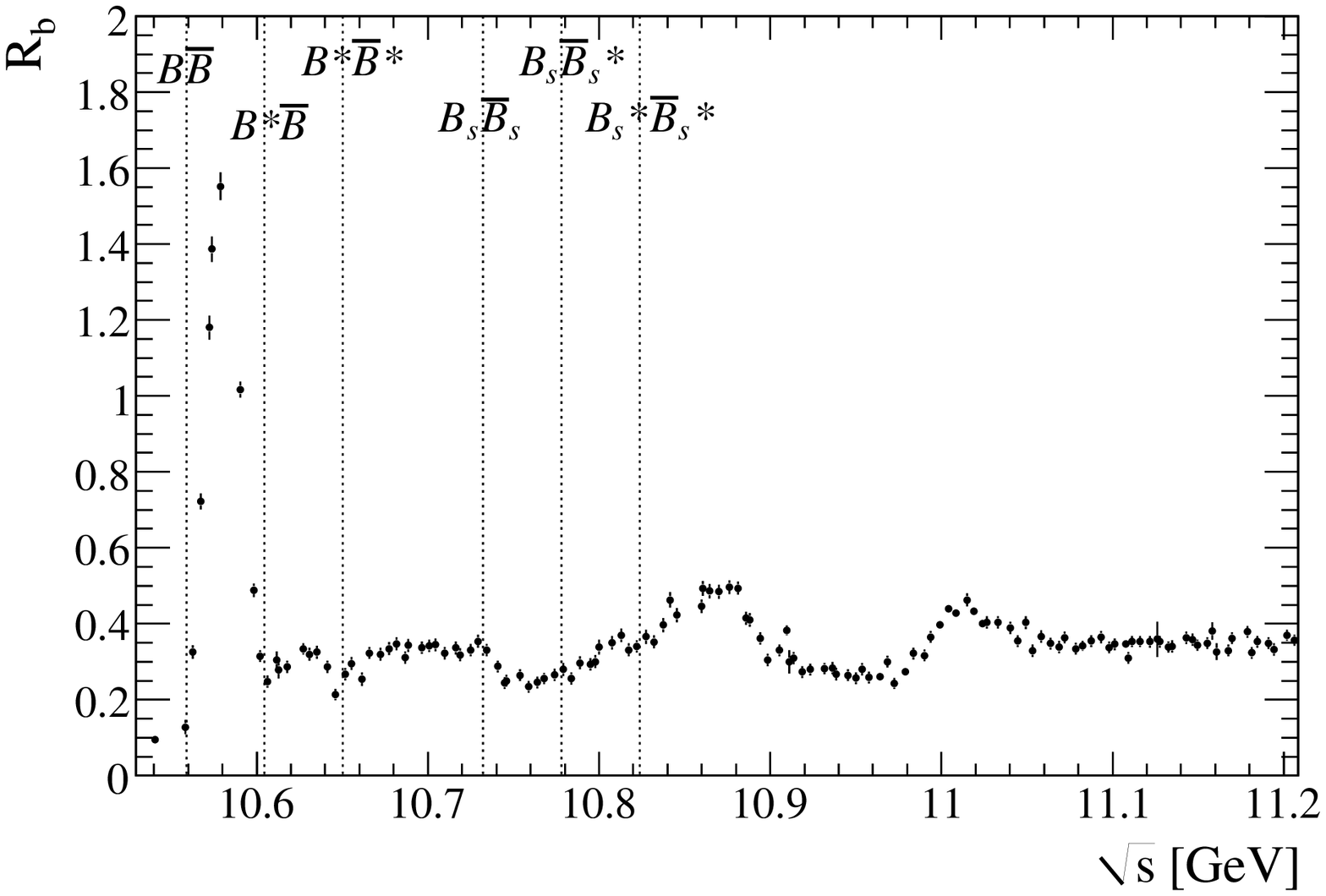,width=\columnwidth}
\epsfig{file=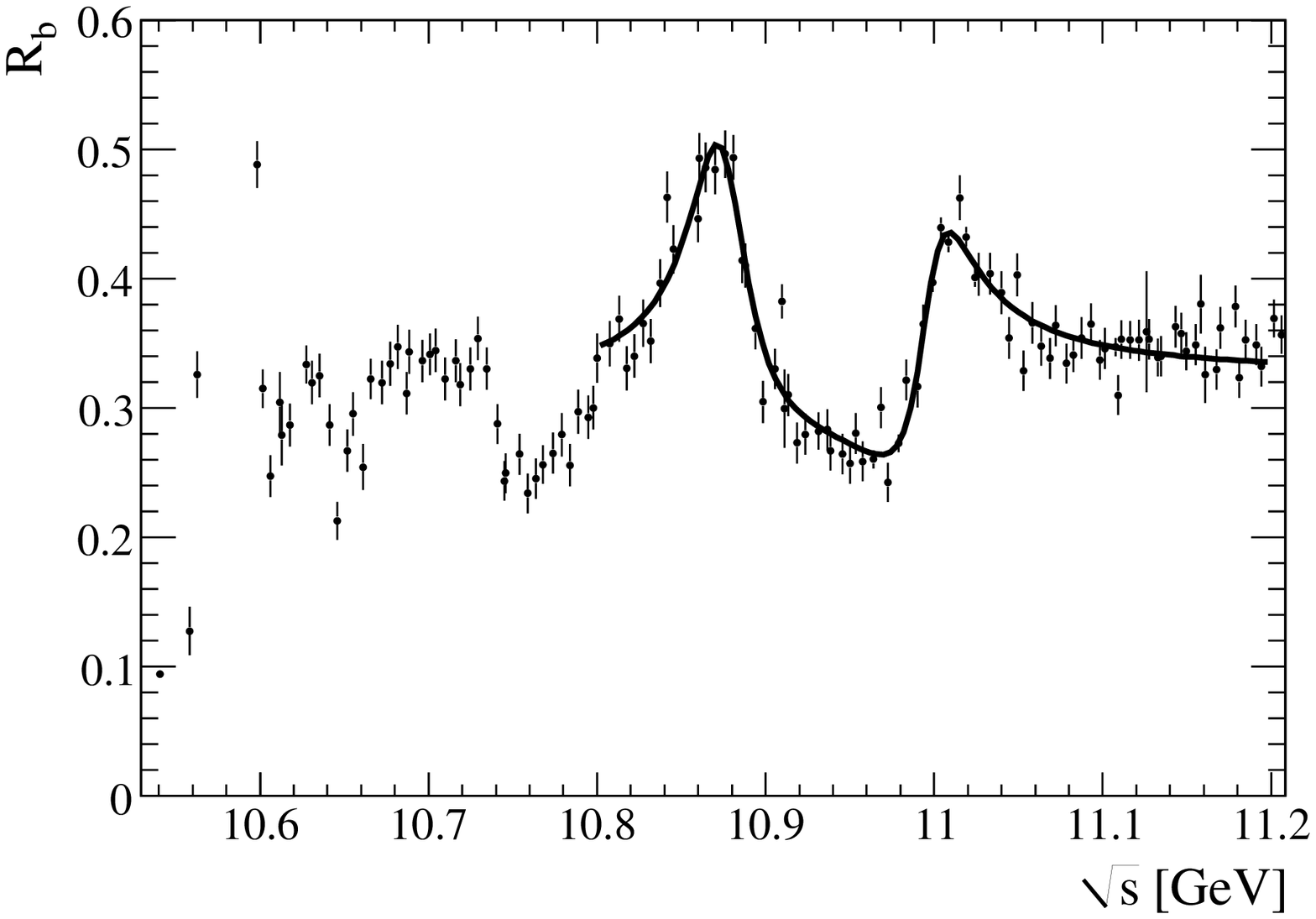,width=\columnwidth}
\end{center}
\caption{(Left) Measured $R_b$ as a function of \SqrtS\ with the position of
the opening thresholds of the $e^+e^-\to B_{(s)}^{(*)}\bar{B}_{(s)}^{(*)}$
processes indicated by dotted lines. (Right) A zoom of the same plot with 
the result of the 
fit
described in the text superimposed. The errors on data represent the 
statistical and the uncorrelated systematic errors added in quadrature.}
\label{fig:results}
\end{figure*}

In the following we describe the method used to derive the inputs
to Eq.~\ref{eq:rb} and the corresponding errors, separating correlated and
uncorrelated errors.
The covariance matrix for the measurements of $R_b$ at
different energies is $V_{ij} = \left[\sigma^2_{stat}(s_i) +
\sigma^2_{unc}(s_i)\right]\delta_{ij} + 
\sigma_{corr}(s_i)\sigma_{corr}(s_j)$, where $\sigma_{stat}(s_i)$,
$\sigma_{corr}(s_i)$, and $\sigma_{unc}(s_i)$ are the
statistical, correlated, and uncorrelated systematic error respectively, and
$\delta_{ij}$ is the Kronecker delta.
 
The efficiency for the di-muon selection $\epsilon_\mu$ 
is extracted from a sample of fully simulated MC events generated with
KK2f~\cite{kk2f} at several values of \SqrtS.
Due to the change in boost
this efficiency is found to change by
 1.5\% over the whole range and the MC statistics error we assign to the 
corresponding correction is
0.2\%. The correlated uncertainty on the absolute scale of the efficiency is
estimated to be 1\% and to come primarily from uncertainties in the 
simulation of the trigger,
of the quantities used in the selection and of the 
tracking efficiency. We also account for differences in
the trigger configurations between the scan data and the reference data taken
during the year 2007  and estimate the efficiency on the 
reference data to be lower by $(0.5\pm0.2)\%$. The same generator is
consistently used to extract $\xi_\mu=1.48\pm0.02$, where this correlated error
is due to the uncertainty on the cross-section. 

The efficiency for $e^+e^-\to b\bar{b}$ events is estimated by using
EvtGen~\cite{evtgen} as generator, separately for each possible two-body final
state including $B$, $B_s$, and $B^*_s$ mesons, and at different
values of \SqrtS.
Because we ignore the relative composition in terms of 
final states at each energy we consider the largest and 
the smallest efficiencies among the allowed final 
states and take their mean value as the central value and half their 
difference  as uncorrelated error.
The correlated error on the absolute scale of $\epsilon_B$ 
is estimated by varying the selection criteria
and it is found to 
amount to 1.3\%.

The calculation of the double ratio $\kappa_{\sigma\epsilon}$ requires  
the  dependence on \SqrtS\
of $\epsilon_\mu$, which has already been discussed,  and the 
cross-sections and efficiencies for the $ISR$ and the background 
processes. 

The $ISR$ cross-section is computed to second-order according to
Ref.~\cite{ISR}. The corresponding efficiency ($\epsilon_{ISR}$) is 
estimated with MC simulation to be 41\% on average. The relative efficiency
change across the scan, estimated to be $\sim 5\%$, is used as a correlated
uncertainty and it propagates to an error on $R_b$ of at most 0.7\%.

The cross-section for two-photon events scales as 
the square of the logarithm of $s$, and 
the corresponding efficiency is considered to be flat. 
The product of the cross-section and the 
efficiency ($\sigma_{\gamma\gamma}\epsilon_{\gamma\gamma}$) before the 
$R_2$ is fitted 
from the distribution of the
direction of the missing momentum and then multiplied by the $R_2$ cut 
efficiency. We attribute 50\% uncertainty to this estimate,  
leading to a relative correlated error of at most 0.2\%. 
Finally,
the product of the continuum cross-section and efficiency is computed by 
subtracting the
$ISR$ and two photon components from $N_h^\prime$ (see Eq.~\ref{eq:nhp}). 
The continuum contribution to $R$ ($R_{cont}$) is assumed to be constant with
\SqrtS, while the corresponding efficiency
($\epsilon_{cont}$) was estimated on a sample of MC events generated with
JETSET~\cite{JETSET}. No correction to account for the fact that the reference
data were taken in a different data-taking period was found necessary. The
relative change of $\epsilon_{cont}$ over the whole scan range is estimated to
be 3\% and a 0.2\% systematic error due to MC statistics is assigned to it. We
also find that the distribution of $R_2$ is not perfectly reproduced 
by the MC. We therefore estimate the scaling of $\epsilon_{cont}$ 
separately with and without the $R_2<0.2$ requirement and take the difference
among the results as a correlated systematic error. Its contribution 
depends on the value of $R_b$ and it is at most 2\%.

To measure \SqrtS\ of each point we fit the distribution of the invariant 
mass of the two muons in the selected di-muon sample with a function made of a 
Gaussian with an exponential tail on the side below the peak
mass. We then use the mean of the Gaussian as
estimator of \SqrtS\ and we determine a bias of $(20.9\pm1.5)\mev$ for this
quantity by comparing the \ThreeS\ mass measured on the data taken during 
the $\sim 100 \invpb$ scan performed by PEP-II at the beginning of the last 
data-taking period  with the resonant depolarization result~\cite{VEPC}. 
We correct for this bias, that comes from the (strongly) non linear impact 
of the momentum resolution in the invariant mass, and verify on simulated 
events that it does not depend on $\SqrtS$.
\begin{table}
\begin{center}
\caption{ Contributions to the relative correlated systematic error on 
$R_b$. \label{tab:sys}}
\begin{tabular}{|l|c|}
\hline
Contribution & Relative error (\%)\\ \hline
$\mu\mu$ MC statistics & 0.2\\
$\mu\mu$ radiative corrections & 1.4\\
$\epsilon_\mu$ & 1.3\\
$\epsilon_B$ & 1.3\\
$\epsilon_{cont}$ &$ <2.0$\\
$\epsilon_{ISR}$ &$ <0.7$\\
$\sigma_{\gamma\gamma}\epsilon_{\gamma\gamma}$ &$ <0.2$\\
\hline
\end{tabular}
\end{center}
\end{table}

\begin{table}[bht]
\begin{center}
\caption{ Fit results for the \yfive\ and \ysix\ resonances resulting from 
the fit described in the text. The $\phi$ phases
are relative to the interfering continuum. The corresponding world 
averages~\cite{PDG} are also reported.\label{tab:resonances}}
\begin{tabular}{|c|c|c|}
\hline
&\yfive&\ysix\\ \hline
 mass (GeV) & $10.876\pm0.002$& $10.996\pm0.002$\\
width (MeV) & $43\pm4$&  $37\pm3$\\ 
$\phi$ (rad) & $2.11\pm0.12$&$0.12\pm0.07$\\ \hline
PDG mass (GeV)& $10.865\pm 0.008 $&$11.019\pm 0.008$\\
PDG width (MeV) &$110\pm13$&$79\pm 16$ \\ \hline
\end{tabular}
\end{center}
\end{table}

The resulting measurements of $R_b$ as a function of \SqrtS\ are shown in 
Fig.~\ref{fig:results}, where the error bars represent the sum 
of the statistical and uncorrelated systematic errors and dotted lines 
show
the different $B$ meson production
thresholds. The relative correlated systematic errors on $R_b$
are summarized in Table~\ref{tab:sys}. The numerical results for each energy
point, together with the estimated ISR cross section, can be found in
Ref.~\cite{epaps}. It is important to stress that radiative corrections have not
been applied since they would require an a-priori knowledge of the resonant
region. The measured $R_b$ therefore includes all final- or initial- state 
radiation 
processes.

The large statistics and the small energy steps of this scan 
make it possible to observe clear structures corresponding to the 
opening of new thresholds: dips corresponding to the $B^{(*)}B^*$ and 
$B_sB_s^*$ openings and a plateau close to the $B_s^*B_s^*$ one.
It is also evident that the \yfive and \ysix behave 
differently above and below the corresponding peaks. Finally, the plateau above the 
\ysix is clearly visible.

We fit the following simple model to our data between
10.80 and 11.20 GeV: a flat component representing $b\bbar$-continuum
states not interfering with resonance decays, added incoherently to a
second flat component interfering with two relativistic Breit Wigner
resonances: $\sigma=|A_{nr}|^2+|A_r+A_{10860}e^{i\phi_{10860}}BW(M_{10860},\Gamma_{10860})+
A_{11020}e^{i\phi_{11020}}BW(M_{11020},\Gamma_{11020})|^2$, with $BW(M,\Gamma)=1/[(s-M^2)+iM\Gamma]$.
  The results summarized in 
Table~\ref{tab:resonances} and Fig.~\ref{fig:results} differ substantially
from the PDG values. In particular the $B_s^*B_s$ and 
$B_s^*B_s^*$ thresholds have a very large impact on the determination of 
the \yfive\ width.

The number of states is, a priori, unknown as are their energy
dependencies. Therefore, a proper coupled channel
approach~\cite{eichten,thornquist} including the effects of
the various thresholds outlined earlier, would be likely to modify the results
obtained from our simple fit.  As an illustration of the systematic
uncertainties arising from the assumptions in our fit, a simple modification is
to replace the flat nonresonant term by a threshold function at $\SqrtS = 2
m_B$. This leads to a larger width ($74\pm4\mev$) and a lower mass
($10869\pm2\mev$) for the \yfive.

In summary, we have performed an accurate measurement of $R_b$ in fine 
grained center-of-mass energy steps and have shown that these measurements 
have the potential to yield information on the bottomonium 
spectrum and possible exotic extensions.

We are grateful for the excellent luminosity and machine conditions
provided by our \pep2\ colleagues,
and for the substantial dedicated effort from
the computing organizations that support \babar.
The collaborating institutions wish to thank
SLAC for its support and kind hospitality.
This work is supported by
DOE
and NSF (USA),
NSERC (Canada),
CEA and
CNRS-IN2P3
(France),
BMBF and DFG
(Germany),
INFN (Italy),
FOM (The Netherlands),
NFR (Norway),
MES (Russia),
MEC (Spain), and
STFC (United Kingdom).
Individuals have received support from the
Marie Curie EIF (European Union) and
the A.~P.~Sloan Foundation.

\end{document}